\documentclass[a4paper,10pt]{article}
\usepackage{graphicx}
\usepackage{amsfonts}
\usepackage{amssymb,amsmath}

\title{ON GRAVITY AND MASS}
\author{Athanasios   Markou\footnote{athanasios.markou@cern.ch}\\  Institute of Nuclear Physics \\ NCSR DEMOKRITOS \\ P.O Box 60228 \\ 15310 Ag. Paraskevi (Attiki), Greece}
\date{November 12, 2008}
\begin{document}

\maketitle

\begin{abstract}
Motivated mainly by the fact that no charged elementary particles having zero mass have been observed up to now, we investigate the question whether the mass of the elementary particles is 
connected with their electric charge and whether gravity can be derived from QED. The graviton is considered as a two-photon bound state. A relation between mass and charge of elementary particles is derived. Masses of the light quarks $d$ and $u$ are calculated from the electron mass, assuming that $d$, $u$ and $e$ are all fundamental and not composite. In this picture, the heavier quarks and leptons are considered as not fundamental, the massive neutral gauge bosons (and then their charged partners)  are composite. The here calculated $u$ and $d$ quark masses, result in quark-mass ratios which display interesting regularities. The lightest quark mass turns out to be quite small. This may be interesting in connection to the strong CP problem.
\end{abstract}

\begin{center}
strong CP problem
\end{center}

\section{Introduction}
According to our present views, in the Standard Model of elementary particles masses are given to fermions and gauge bosons by the Higgs mechanism. However, this mechanism is rather a parametrization, does not provide a deeper insight into the problem of the origin of mass, and can not explain the hierarchy of the fermion masses. In classical Physics, General Relativity attributes the curvature of spacetime to the existence of mass, energy and momentum.  On the other side, as W. Pauli writes in his “"Theory of Relativity"”, General Relativity "“does not provide a physical interpretation for the sign (gravitational attraction and not repulsion) and numerical value of the gravitational constant, but takes these data from experiment”".
Obviously, some important Physics is hidden in the numerical constant and the sign. We believe that these problems are interconnected, and this work presents an effort to approach them as a whole.
\section{Gravity and mass from QED ?}

It is an experimental fact that no charged elementary particle with zero mass, has been observed so far.  On the other hand, all existing neutral particles with the possible exception 
of the $Z^{0}$ boson, are known to be either composed of charged constituents and are massive,  or they are elementary and massless (or have a tiny mass). For the $Z^{0}$ boson, a number of publications suggest that it may be composite.  Based on these observations and excluding for the moment the $Z^{0}$ from our reasoning, we make the following working hypothesis:  The mass of the elementary particles is (mainly) of electromagnetic origin and gravity is (mainly) 
an electromagnetic effect. Such an approach, although it originated very early  (based on different reasoning)\cite{1a,1b,1c}, may irritate some readers, but one has to take into account that  in the last 3-4 decades too, a rather extensive literature on the “electromagnetic mass” of elementary particles has appeared. 
Within the framework of General Relativity it has been stated \cite{1d} that “…of the energy constituting matter, three quarters 
is to be ascribed to the electromagnetic field and one-quarter to the gravitational field”. In addition, considerable literature exists, 
on “{\em induced}” or “{\em emergent}” gravity, and on the question whether gravitation is a fundamental interaction, or if General Relativity is a long range 
effective theory. Finally we should note that eminent physicists have worked on the question of the “electromagnetic mass” and mention Feynman's statement \cite{1e}, that there is "the thrilling possibility~... that the mass is all electromagnetic".

We will thus try to investigate whether the gravitational potential can be deduced from electromagnetism, or more precisely from the QED.  It is known, that potentials make more sense in the static regime and classically they have to be put in by hand, but they can be computed in QED and QCD from the exchange of field quanta in particle scattering \cite{2,3}.

The calculation of the particle exchange interaction potentials perturbatively in  Quantum Field Theory from particle exchange, has another important advantage over Classical Theory: It allows us to determine the sign of the interaction (attraction or repulsion),  through the spin of the exchanged particles \cite{3,3b,4,5,6}:  For particles with the same “charge”, even-spin exchange corresponds to attraction and odd-spin to repulsion.

Presently, gravity is thought to be mediated by the hypothetical massless spin-2 graviton. From the considerations above and since we have made in the beginning the working hypothesis that gravity may be (mainly) an electromagnetic effect, we are led to the assumption that the graviton may be composed of two photons. A two-photon state could have spin 0,1 and 2. A composite graviton is not something new. In the literature it is already indicated that the graviton may be composite.

An analogous situation exists in nucleon-nucleon scattering which proceeds via (the composite) pion exchange, or in hadron scattering, which at intermediate 
energies is thought to be due to correlated two-pion exchange and is described in terms of rho and sigma meson exchanges   \cite{7,8,9}. In the linear sigma model, a phenomenological model of the mutual two-pion interaction has been constructed, supporting the correlated two-pion exchange mechanism in N-N interactions \cite{10,11}. Correlated $q\bar{q}$ exchange is also believed to exist in effective theories of QCD too \cite{12} .

Photons can interact with other photons through virtual electron-positron pairs created in vacuum (vacuum polarization). The existence of photon-photon interactions was first proposed in 1934 \cite{13a}  and was theoretically discovered  shortly afterward \cite{13b,13c}. Since then, further QED calculations on this subject have been carried out \cite{13d,13e}. At high energies, interactions between virtual photons are well established, and interactions between real photons have been relatively recently observed \cite{13f}. At low energies the cross section is extremely small, of the order of $10^{-65} cm^2$ at 1 eV \cite{13e,14a}. Therefore photon-photon interactions at low energy are very difficult to observe. Only upper limits ($1.5\times10^{-48} cm^{2}$) exist up to date \cite{14a}, but efforts are being made to detect such interactions \cite{14b}.

 In versions of Abelian theories defined in non-commutative spaces (interest on such theories has grown considerably recently), photons show a self-interacting behavior  like the one known from QCD, so that photon-photon bound states (spin 0,1 and 2) are possible \cite{15}. Two-photon bound states are thought to be formed and are being searched for in atomic/optical   physics experiments \cite{16a,16b,16c}.

 If  low energy photon-photon interactions would be attractive, they could form a graviton as a two-photon bound state in the framework of QED.  Indeed \cite{17a}, the amplitudes for low energy photon-photon scattering are negative if the outgoing photons have the same polarization \cite{17a}. Therefore, since the potential of the interaction is the 
Fourier transform of the amplitude, the graviton as a two-photon bound state would be possible
 only for spin 2 and not for spin 0 or 1. The potential of the photon-photon interaction is negative also at the two-fermion-loop level for photons with the same 
polarization \cite{17b}. In  \cite{16d} the possibility of photon-photon interactions giving rise to gravitons has been investigated within the framework of QED 
\footnote{S. Guttenberg brought to our attention this reference, after reading a late version of the present article.}.

 Based on the above considerations regarding possible photon-photon bound states, we are guided to consider the graviton as a two-photon bound state with spin-2  and  gravity at the quantum level as the result of the exchange of such a state.

In QFT, the exchange of a single particle between two scattering particles, is known  to result in the Yukawa  potential\cite{3} , which was developed originally for the case of the pion exchange,  but it is valid more generally and is used even in the case of gravity, as one can see from the PDG \cite{PDG}, where graviton mass deviations from the value of zero are given 
assuming a Yukawa potential. For massless particle exchange between charged particles the Yukawa potential goes over to a Coulomb-like potential with an $1/r$ behavior.

From the above, as in the case of the pion (a quark-antiquark composite) exchange, we believe that the exchange of a massless graviton, considered to be a two-photon bound state, would result in a $1/r$ behavior.  This should be particularly noted, since it is known that the exchange of two or more uncorrelated photons, does not lead to a $1/r$ dependence.

Fig.~\ref{fig:amtEPS} shows a diagram for the exchange of a two-photon composite graviton between two charged elementary particles $1$ and $2$ with charges $Q_1$ and $Q_2$.

\begin{figure}
\label{fig:amtEPS}
   \centering
   \includegraphics[scale=1]{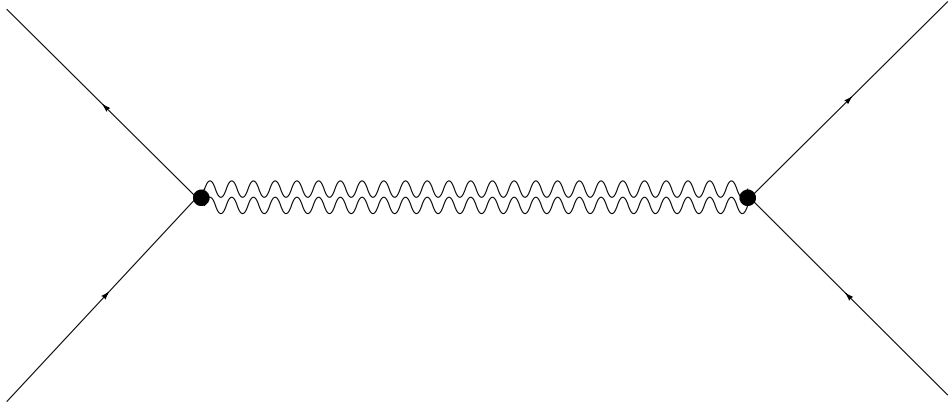}
   \caption{Exchange of a two-photon composite graviton between two charged elementary particles .}
   \label{fig:amtEPS}
\end{figure}

If the graviton were  a two-photon composite, two photons would be coupling at each of the two vertices in Fig.~\ref{fig:amtEPS}, giving a potential proportional to $Q_1^2$ times $Q_2^2$ ,
\begin{equation}
\label{eq1}
V = -K\frac{Q_1^{2}Q_2^{2}}{r}
\end{equation}
where the minus sign would be due to the (even) spin of the exchanged particle.  From this relation we see that the interaction would be attractive, independent of the sign of $Q_1$ and $Q_2$.

The constant K would include the details of the exchange of the two-photon bound state. In order to calculate the scattering amplitude, the product of the two-photon bound state propagator and the amplitudes for emission of such states by the two charged particles is needed. 
Since at low energies the photon-photon interaction is extremely weak (the cross section is of the order of $10^{-65} cm^{2}$ at 1 eV), it may be possible that the two-photon bound state amplitudes 
have the appropriate value to give an attraction consistent with the strength of gravity. 

One should derive Eq.~\ref{eq1} from QED and calculate the constant K. An empirical estimate for K is given below.

          If these ideas would be correct, the Coulomb law would have to be modified to 
\begin{equation}
F = K_e \frac{Q_1 Q_2}{r^2} - K\frac{Q_1^2 Q_2^2}{r^2}
\end{equation}
and the following equality  would exist 
\begin{equation}
 K\frac{Q_1^2Q_2^2}{r^2}=G_N\frac{m_{1}m_{2}}{r^2}\text{.}
\end{equation}

 Therefore for the gravitational interaction between identical and charged, elementary\footnote{As charged elementary particles, we consider only particles that are not composite (to our present  knowledge) and are stable, i.e the d and u quarks and the electron. Since
     the heavier quarks and leptons are  not stable, we consider them as  excited states, and
     their masses as not fundamental quantities. The idea that some of the quarks could be elementary and others composite, can be found in the literature \cite{35c}\cite{36c}\cite{37c} too.  In order to calculate the masses of composite objects, one would have to add the
contributions coming from all charges of the constituents (including those of the
virtual sea-quark pairs), to the contributions of all other interactions besides QED.}
     particles   we would have 
\begin{equation}
\label{eq2}
K\frac{Q^{4}}{r^2}=G_N\frac{m^2}{r^2} \text{ or } m =Q^2\sqrt{\frac{K}{G_N}}  \text{ or }  K=G_N\frac{m^2}{Q^4}\text{.}
\end{equation}

\section{Particle masses and quark mass ratios}
	Unlike in classical physics,  Eq.~\ref{eq2} connects the mass and charge of particles, without using a particle radius and  allows the determination of $K$ empirically, 
rom the known mass and charge of a massive charged elementary particle. 
If we would know  that  the electron is indeed elementary and not composite  Eq.~\ref{eq2}  would  result in

\begin{equation}
\label{eq3}
  K = 0.0851 [m^3 kg s^{-2} C^{-4}] .
\end{equation}

 From this value and  Eq.~\ref{eq2}, the  masses of the elementary quarks d and u could be calculated to first order. Our calculation gives a mass of 0.23 MeV for the $u$ quark ($Q=2/3 e$), and a mass of 0.057 MeV for  the $d$ quark. Thus $m_u/m_d=4$ and $m_u >m_d$ .  It is presently widely accepted that the opposite is true, but one has to consider that the light quark masses can not be measured and the currently accepted relation $m_u<m_d$   is based  mainly on the observation  that  hadrons containing  $d$ quarks have  larger masses than those containing u-quarks. However hadron masses are of dynamical origin, the underlying effects are non-perturbative (and unknown to a large extent), and therefore this derivation may not be true. Furhtermore,  $m_u <m_d$   means  that this would be  the only (and somewhat strange) exception to the observation  that the quarks with larger charge have also larger mass. Some doubts on the validity of the widely accepted inequality $m_u <m_d$ have been published \cite{19a,19b,19c} and in some cases $m_u >m_d$ has been derived \cite{19d,19e, 19f}. 
Furthermore, publication \cite{last} derives the ratio $m_u/m_d$ as a function of mass ratios of the heavier quarks $s$, $c$, $b$, and $t$. If we use their relation and the most recent PDG values 
for the masses of the $s$, $c$, $b$ and $t$ quarks, we get $m_u >m_d$ $(m_u/m_d  =2.6)$. If we use the slightly smaller mass of $80 MeV$ for the $s$ quark (well within the $25 MeV$ error given by the PDG), but keep the other quark masses as before since they are known to much better precision, we arrive approximately at the value $m_u/m_d =4$, given above from our calculation of the u and quark masses from their charge.

      On the other hand,  if we use the  u- and d- quark masses calculated above and the masses of the other quarks as given in the PDG tables, we get the following  (approximate) quark-mass ratios for the up- type quarks: $m_c /m_u =5434$ and  $m_t /m_c =139$. For the down- type quarks we obtain:$m_s /m_d =1667$,  $m_b /m_s =47$. These ratios display some interesting regularities, in contrast to the ratios obtained  when using the generally accepted u and d quark mass values. The most obvious regularity is that the mass ratio  of the second quark to the first is much larger than the ratio of the  
third to the 
second, within each Q=2/3e and Q=1/3e group. (Interestingly the same is true for the ratio of masses muon/electron and tau/muon). Such  regularities  may be expected if there were  a common underlying dynamics of  the u- and d- type groups.

From  Eq.~\ref{eq2} an upper limit for the photon mass can be calculated using the present experimental limit on its charge:  From  $Q< 5\times10 ^{-30} e$,   we would get  $m< 10^{-52}$ eV. This is many orders of magnitude beyond the present experimental limit. 

\section{Discussion}
Experimental tests of these ideas are difficult. If these ideas are correct, the $Z^{0}$ (and therefore also the $W^{+}$, $W^{-}$) and the Higgs particle(s) would be composite. This could be investigated in accelerator experiments. A definitive test of the masses of the $u$ and $d$ quarks predicted here, does not seem possible since -as we presently believe- quarks are permanently confined inside hadrons. 
Therefore they cannot be observed as free particles, so that their masses cannot be measured. It may be possible though to have an estimate of the 
$u$, $d$ masses if we look at hadron properties which depend on the current quark masses, instead of looking at hadron masses usually calculated in constituent quark models. Such properties have been calculated recently in \cite{20}, but for equal masses of the $u$ and $d$ quarks. It has been suggested \cite{last1, last2}, that if one of the quarks is massless (or has a very small mass), this may mean a solution to the strong CP problem. Since the here calculated mass of the lightest quark is quite small, it may be interesting to investigate if this could help in connection to the strong CP problem. In this work, the graviton is considered as a two-photon bound state. This would mean that the graviton may have an (extremely small) mass, leaving room for a deviation from the $1/r^2$ dependence at very large distances. 

\section{Acknowledgments}
    We thank S. Guttenberg, C.G. Papadopoulos, A. Petkou  for useful discussions and  comments.

\end{document}